\onecolumn  
\def\@oddhead{}\def\@evenhead{}
\def\@oddfoot{\rm\rightmark \hfil Page \thepage}
\def\@evenfoot{\@oddfoot}

\def\maketitle{\par
  \begingroup
  \def\thefootnote{\fnsymbol{footnote}}
  \def\@makefnmark{\hbox
  to 0pt{$^{\@thefnmark}$\hss}}
  \onecolumn\@maketitle  
  \@thanks
  \endgroup
  \setcounter{footnote}{0}
  \let\maketitle\relax
  \let\@maketitle\relax
  \gdef\@thanks{}\gdef\@author{}\gdef\@title{}\let\thanks\relax}
\def\@maketitle{\vbox to 3in{\hsize\textwidth
  \linewidth\hsize
  \vfil
  \begin{center}{\large\bf \@title \par}\end{center} \vskip 1em   
  {\normalsize\begin{center}\begin{tabular}[t]{c}\@author
  \end{tabular}\end{center}\par}
  \vfil}}

\def\onehead#1#2{\vskip0.5cm\noindent\hbox to 3pc{\bf #1\hfil}{\bf #2}
                 \vskip0.3cm}
\def\twohead#1#2{\vskip0.5cm\noindent\hbox to 3pc{\normalsize\it
#1\hfil}{\it #2}\vskip0.3cm}
\def\ttwohead#1#2{\vskip0.2cm\noindent\hbox to 3pc{\normalsize\it
#1\hfil}{\it #2}\vskip0.3cm}

\def\Onehead#1#2#3{\vskip0.5cm\noindent\hbox to 3pc{\bf #1\hfil}%
                 {\bf #2}\newline
                 \hbox to 3pc{\hfil}{\bf #3}\vskip0.3cm}

\def\thebibliography#1#2{
  {\onehead{\bf #1}{REFERENCES}
  \@mkboth {REFERENCES}{REFERENCES}}\list
  {\arabic{enumi}.}{\settowidth\labelwidth{[#2]}\leftmargin\labelwidth
  \advance\leftmargin\labelsep\itemsep0pt \frenchspacing
  \usecounter{enumi}}
  \def\newblock{\hskip .11em plus .33em minus -.07em}
  \sloppy
  \sfcode`\.=1000\relax}

\documentstyle[B_Phys,12pt]{article}
\begin{document}
\begin{flushright}
FSU--HEP--930719\\
July 1993
\end{flushright}
\vglue 0.5cm
\begin{center}
{\large\bf PROBING THE $WW\gamma$ VERTEX IN RADIATIVE $b$-QUARK \\
DECAYS}
\footnote{To appear in the Proceedings of the {\it ``Workshop on $B$
Physics at Hadron Accelerators''}, Snowmass, CO, June~21 --~July~2, 1993.}
\vglue 0.6cm
{U.~BAUR\\
        \em Physics Department, Florida State University, \\
        \em Tallahassee, FL 32306, USA}
\end{center}
\vglue 1.cm
 \pagestyle{plain}    
\centerline{\bf Abstract}
\vglue 0.2cm
{\rightskip=3pc
 \leftskip=3pc
 \baselineskip=14pt
 \noindent
The recent CLEO results on on radiative $b$-quark decays are used to derive
constraints on anomalous $WW\gamma$ couplings. These constraints are
compared with expectations from $p\bar p\rightarrow e^\pm
p\llap/_T\gamma+X$ at the Tevatron. The usefulness of exclusive
radiative $B$ meson decay channels in probing the $WW\gamma$ vertex is
largely limited by present theoretical uncertainties in the calculation
of hadronic matrix elements.
\vglue 0.6cm}

One of the major goals of future experiments at the Tevatron is to probe
the structure of the $WW\gamma$ vertex in $W\gamma$ and $W^+W^-$
production. Such direct tests of three vector boson vertices through
tree level processes have to be contrasted with indirect tests which
involve one-loop processes. Whereas bounds derived from tree level
processes are essentially model independent, limits on anomalous
$WW\gamma$ couplings extracted from processes which are sensitive to
three vector boson couplings only at the one-loop level usually do
depend on specific assumptions \cite{LOOP}. The dependence on model
specific assumptions is most pronounced in quantities where anomalous
couplings lead to divergencies, {\it e.g.} the $S$, $T$ and $U$ parameters.

Some one-loop processes, such as $b\rightarrow s\gamma$, yield finite
answers due to the GIM mechanism. Recently, the CLEO Collaboration
reported \cite{CLEO} the observation of the decay $B\rightarrow
K^*\gamma$ with a branching fraction of $B(B\rightarrow K^*\gamma)=
(4.5\pm 1.5\pm 0.9)\cdot 10^{-5}$. In the following we analyze the
implications of this measurement on the anomalous $WW\gamma$ couplings,
$\Delta\kappa$ and $\lambda$, and compare the result with expectations
from future experiments at the Tevatron.

Our calculations are based on the results obtained in Ref.~\cite{ANOM}
for the inclusive radiative decay $b\rightarrow s\gamma$ for arbitrary
anomalous couplings $\Delta\kappa$ and $\lambda$. Apart from
non-standard
contributions to the $WW\gamma$ vertex we assume the Standard Model to
be valid. QCD corrections are
incorporated following Ref.~\cite{ROX}. To estimate the branching
fraction of the exclusive decay mode $B\rightarrow K^*\gamma$ we use the
approach of Ref.~\cite{ALI}. In this model, $B(B\rightarrow K^*\gamma)$
is estimated by integrating the invariant mass distribution of the
hadrons recoiling against the photon from the $m_K+m_\pi$ threshold up
to ${\cal O}(1$~GeV), assuming that this range is completely saturated
by the $K^*$ resonance. The upper integration limit, however, is only
loosely defined. Together with uncertainties in the $B$ meson wave
function, this results in rather large uncertainties in the estimated
$B\rightarrow K^*\gamma$ branching ratio. For the present lower
experimental limit on the top quark mass \cite{TOP}, $m_{\rm
top}=108$~GeV, we find
$B(B\rightarrow K^*\gamma)=(2-9)\cdot 10^{-5}$. For $m_{\rm
top}=200$~GeV, we obtain $B(B\rightarrow K^*\gamma)=(3-12)\cdot 10^{-5}$.
These ranges are consistent with the branching ratios obtained in other
models \cite{OTHER}.

The resulting constraints on $\Delta\kappa$ and $\lambda$ depend
explicitly on $m_{\rm top}$, and are shown in Fig.~1.
\begin{figure}[t]
\vskip 7.5cm
\includegraphics{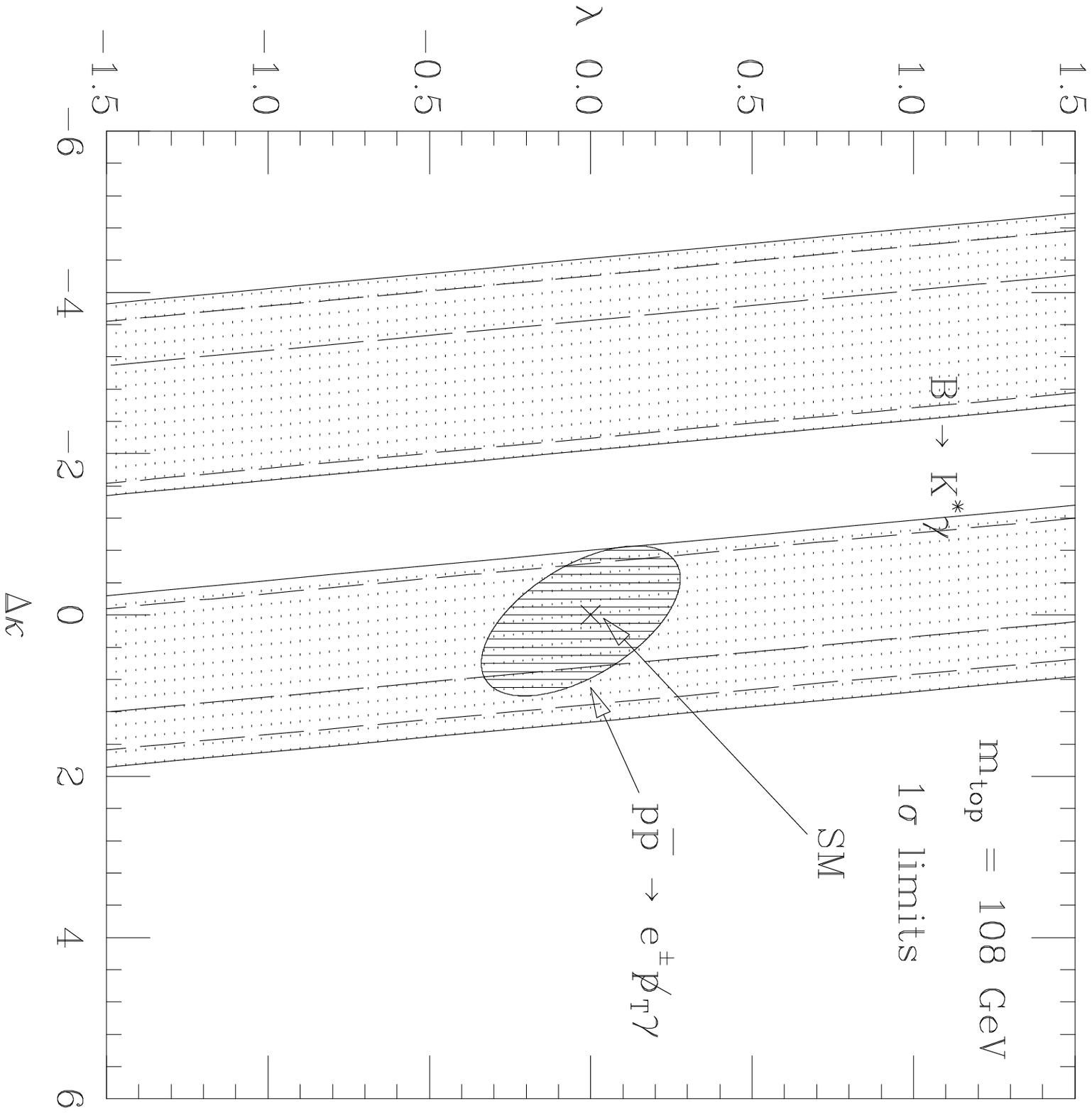}
\includegraphics{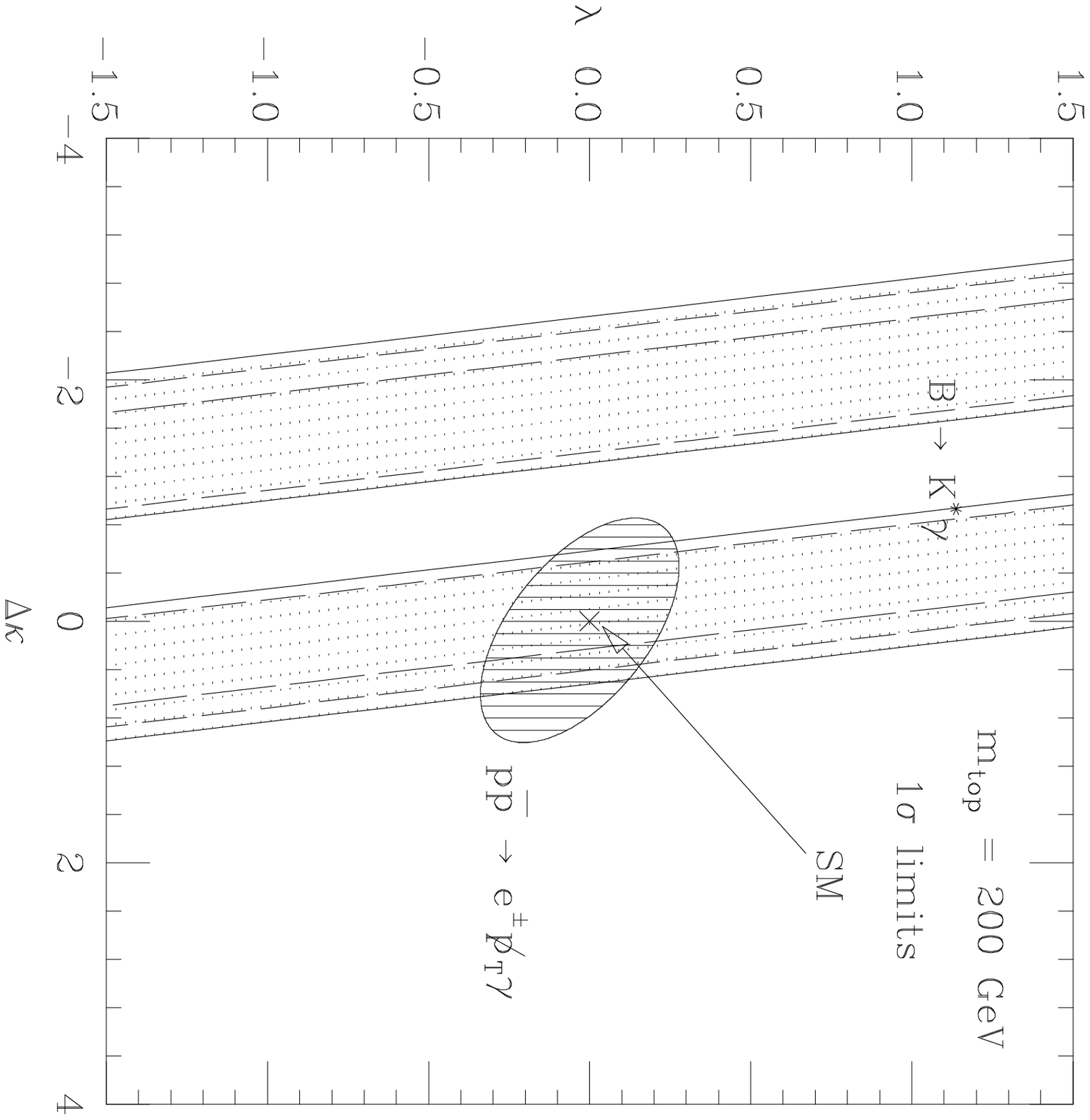}
Figure~1: Allowed regions in the $\Delta\kappa -\lambda$ plane for
$m_{\rm top}=108$~GeV and $m_{\rm top}=200$~GeV. The
region allowed by present $B\rightarrow K^*\gamma$ data is indicated by
the shaded bands. The short-dashed lines outline the limits
from $B\rightarrow K^*\gamma$ expected from CDF with an integrated
luminosity of 100~pb$^{-1}$. The long-dashed lines show the bounds from
the CLEO upper limit on the branching ratio for the inclusive decay
$b\rightarrow s\gamma$. The hatched area,
finally, displays the allowed region in the $\Delta\kappa -\lambda$ plane which
is expected to result from $p\bar p\rightarrow e^\pm p\llap/_T\gamma+X$
at the Tevatron with $\int\!{\cal L}dt=100$~pb$^{-1}$.
\end{figure}
In order to obtain $1\sigma$ limits from $B\rightarrow K^*\gamma$, we
have added the statistical and systematic errors in the branching ratio
linearly. Despite the large uncertainties in the calculation of the
$B\rightarrow K^*\gamma$ decay rate, the CLEO measurement excludes large
regions of the $\Delta\kappa - \lambda$ plane. At the $1\sigma$ level,
only two rather narrow bands remain allowed. The width of these bands
depends quite strongly on $m_{\rm top}$. The region between the two
bands is not excluded with a very high significance; it still allowed at
the $2\sigma$ level.

The CLEO collaboration recently also presented a new upper 95\% CL limit on
the branching ratio of the inclusive decay $b\rightarrow s\gamma$
\cite{APS} of $B(b\rightarrow s\gamma)<5.4\cdot 10^{-4}$, derived from
the inclusive photon energy spectrum in $B$ decays. The
$b\rightarrow s\gamma$ decay rate is much more accurately predicted
theoretically than
that of the exclusive channel $B\rightarrow K^*\gamma$. The region in
the $\Delta\kappa - \lambda$ plane which is consistent with the CLEO
limit on $b\rightarrow s\gamma$ is the one between the two long-dashed
lines in Fig.~1. The bounds obtained from inclusive radiative $b$ decays
reduce the region allowed by $B\rightarrow K^*\gamma$ somewhat. Similar
results have also been obtained in Ref.~\cite{TOM}.

The current measurement of the $B\rightarrow K^*\gamma$ branching
fraction is based on 13~signal events~\cite{CLEO}. A much larger event
sample is possible in the near future from CLEO and, with a special
photon trigger~\cite{CDF}, also from CDF. If this trigger were
implemented, up to 100 $B\rightarrow K^*\gamma$ events are expected in
the 1993-94 run. Assuming that the central value of the branching ratio
does not change, and systematic errors coincide with those of
Ref.~\cite{CLEO}, the anticipated improvement is shown in Fig.~1 by the
short-dashed lines. Since theoretical uncertainties dominate, the
resulting bounds are only slightly better than those obtained with the
present data. A substantial improvement in the calculation of
$B(B\rightarrow K^*\gamma)$, however, may result from a lattice
computation of the hadronic matrix element in the near future. The CDF
photon trigger may
also allow the observation of radiative $B_s$ decays in the channel
$B_s\rightarrow\phi\gamma$ \cite{CDF}. The number of events expected is similar
to the rate foreseen for $B\rightarrow K^*\gamma$. So far, no
theoretical calculation of the $B_s\rightarrow\phi\gamma$ branching ratio
has been performed.

To contrast the bounds on $\Delta\kappa$ and $\lambda$ from radiative
$B$ decays with those from diboson
production, we have also included the $1\sigma$ limits expected from
$p\bar p\rightarrow W^\pm\gamma+X\rightarrow e^\pm p\llap/_T\gamma+X$
with 100~pb$^{-1}$ \cite{BHO} in Fig.~1. $W\gamma$ production is
expected to yield much stronger bounds for $\lambda$ while $B\rightarrow
K^*\gamma$ tends to give better limits for $\Delta\kappa$. The two processes
thus complement each other.

In conclusion, we have shown that present CLEO data on radiative $b$ decays
yield valuable information on anomalous $WW\gamma$ couplings. Future
improvements of limits extracted from exclusive $B$ (and $B_s$) decays
depend mostly on the ability to obtain more accurate estimates of the
hadronic $B$ decay matrix elements. Combined with limits expected from $p\bar
p\rightarrow W\gamma$, $\Delta\kappa$ and $\lambda$ can be highly
constrained in the near future.

\onehead{}{ACKNOWLEDGEMENTS}

I would like to thank A.~Kronfeld, T.~LeCompte, R.~Springer and
J.~Thaler for useful
discussions. This research was supported by the U.S.~Department of
Energy under Contract No. DE-FG05-87ER40319.

\end{document}